\documentclass[twocolumn,aps,showpacs,amsmath,superscriptaddress]{revtex4-1}

\usepackage{bm}
\usepackage{amsmath}
\usepackage{amssymb}
\usepackage[usenames,dvipsnames]{color}
\usepackage{graphicx}
\usepackage{dcolumn}
\usepackage{simplewick}
\usepackage{hyperref}
\usepackage{lineno}

\begin{document}

\title{Electric dipole polarizabilities of alkali metal ions from 
       perturbed relativistic coupled-cluster theory}
\author{S. Chattopadhyay}
\affiliation{Physical Research Laboratory,
             Ahmedabad-380009, Gujarat, 
             India}
\author{B. K. Mani}
\affiliation{Department of Physics, University of South Florida, Tampa,
             Florida 33620, USA}
\author{D. Angom}
\affiliation{Physical Research Laboratory,
             Ahmedabad-380009, Gujarat,
             India}

\begin{abstract}
   We use the perturbed relativistic coupled-cluster theory to compute
   the static electric dipole polarizabilities of the singly ionized
   alkali atoms, namely, Na$^+$, K$^+$, Rb$^+$, Cs$^+$ and Fr$^+$. The
   computations use the Dirac-Coulomb-Breit Hamiltonian with the 
   no-virual-pair approximation and we also estimate the correction to the 
   static electric dipole polarizability arising from the Breit interaction. 
\end{abstract}



\maketitle


\section{Introduction}

 The electric dipole polarizabilities, $\alpha$, of ions are important to 
determine the optical properties of ionic crystals. In addition, for 
closed-shell ions like the singly ionized alkali atoms, $\alpha$ is
a measure of the core-polarization effects in the neutral species. It is, 
however, nontrivial to measure $\alpha$ of ions. For the singly charged alkali 
ions, an indirect method to determine $\alpha$ is through the measurement of 
the transition energy between the non-penetrating Rydberg states of the neutral 
species \cite{mayer-33} and it has been used to determine the $\alpha$ of 
Cs$^+$ \cite{ruff-80,safinya-80}. In absence of experimental data, there is a 
need for accurate theoretical calculations. In the case of neutral atoms, 
accurate values of polarizabilities are essential in studies related to the 
parity non-conservation in atoms \cite{khriplovich-91}, optical atomic 
clocks \cite{udem-02,diddams-04} and physics with the condensates of dilute 
atomic gases \cite{anderson-95, bradley-95, davis-95} are of current interest.

  Theoretically, methods based on a wide range of atomic many-body theories 
have been used to calculate $\alpha$. In this regard, the recent review 
\cite{mitroy-10} provides description about the various theoretical methods 
used to calculate $\alpha$. In the present work we use the perturbed 
relativistic coupled-cluster (PRCC) theory, which was earlier applied
the noble gas atoms \cite{chattopadhyay-12a,chattopadhyay-12b}, to compute the
$\alpha$ of the singly charged alkali ions. The PRCC theory is an extension of 
the standard relativistic coupled-cluster (RCC) theory to include an additional 
perturbation and for this, we introduce a new set of cluster operators. The 
formulation is, however, general enough to incorporate any perturbation 
Hamiltonian. It must be emphasized that, compared to other many-body methods,
the use of PRCC is an attractive option as it is based on coupled-cluster 
theory (CCT) \cite{coester-58, coester-60}: an all order many-body theory 
considered to be reliable and powerful. The recent review \cite{bartlett-07}
provides an overview of CCT, and variants of CCT developed for structure and 
properties calculations. The theory has been widely used for atomic 
\cite{mani-09,nataraj-08,pal-07,geetha-01}, molecular \cite{isaev-04}, 
nuclear \cite{hagen-08} and condensed matter physics \cite{bishop-09} 
calculations. Coming back to the PRCC theory, it is different from the 
other RCC based theories in a number of ways, but the most important one 
is the representation of the cluster operators. In the PRCC theory, the cluster
operators can be scalar or rank one tensor operators and it is decided based
on the nature of the perturbation in the electronic sector. Consequently, the 
theory is suitable to incorporate multiple perturbations of different ranks 
in the electronic sector. 

  One basic advantage of PRCC theory is, it does away with the summation 
over intermediate states in the first order time-independent perturbation 
theory. The summation is subsumed in the perturbed cluster amplitudes and 
this offers significant advantages in computing properties like $\alpha$ 
which involves summation over a complete set of intermediate states.

   The paper is organized as follows. In the Sec. II, for completeness and
easy reference we briefly describe the RCC and PRCC theories with the Breit 
interaction. In Sec. IV we introduce the formal expression of the dipole
polarizability and its representation in the PRCC theory. In the subsequent
sections we describe the calculational part, and present the results and
discussions. We then end with conclusions. All the results presented in this
work and related calculations are in atomic units
( $\hbar=m_e=e=4\pi\epsilon_0=1$). In this system of units the velocity of
light is $\alpha ^{-1}$, the inverse of fine structure constant. For which we
use the value of $\alpha ^{-1} = 137.035\;999\;074$ \cite{codata-10}.


\section{Overview of the coupled-cluster theory}

  The detailed description of the RCC and PRCC theories are given in our
previous works. However, for completeness and easy reference we provide
a brief overview in this section.


\subsection{RCC theory}
  The Dirac-Coulomb-Breit Hamiltonian, denoted by $H^{\rm DCB}$,  is an 
appropriate choice to include the relativistic effects in the structure and
property calculations of high-$Z$ atoms and ions. There are, however, 
complications associated with the negative energy continuum states of 
$H^{\rm DCB}$. These lead to variational collapse and 
{\em continuum dissolution} \cite{brown-51}. One remedy to avoid these 
complications is to use the no-virtual pair approximation. In this 
approximation, for a singly charged ion of $N$ electrons \cite{sucher-80}
\begin{eqnarray}
   H^{\rm DCB} & = & \Lambda_{++}\sum_{i=1}^N \left [c\bm{\alpha}_i \cdot 
        \mathbf{p}_i + (\beta_i -1)c^2 - V_{N+1}(r_i) \right ] 
                       \nonumber \\
    & & + \sum_{i<j}\left [ \frac{1}{r_{ij}}  + g^{\rm B}(r_{ij}) \right ]
        \Lambda_{++},
\end{eqnarray}
where $\alpha$ and $\beta$ are the Dirac matrices, $\Lambda_{++}$ is an 
operator which projects to the positive energy solutions and $V_{N}(r_{i})$ is 
the electrostatic potential arising from the $Z=(N+1)$ nucleus. 
Projecting the Hamiltonian with $\Lambda_{++}$ ensures 
that the effects of the negative energy continuum  states are removed from the 
calculations. The last two terms in $H^{\rm DCB}$, $1/r_{ij} $ and 
$g^{\rm B}(r_{ij})$,  are the Coulomb and Breit interactions, respectively.  
The later, Breit interaction, represents the transverse photon interaction and 
is given by
\begin{equation}
  g^{\rm B}(r_{12})= -\frac{1}{2r_{12}} \left [ \bm{\alpha}_1\cdot\bm{\alpha}_2
               + \frac{(\bm{\alpha_1}\cdot \mathbf{r}_{12})
               (\bm{\alpha_2}\cdot\mathbf{r}_{12})}{r_{12}^2}\right].
\end{equation}
The Hamiltonian satisfies the eigen-value equation
\begin{equation}
   H^{\rm DCB}|\Psi_{i}\rangle = E_{i}|\Psi_{i}\rangle , 
\end{equation}
where, $|\Psi_{i}\rangle$ is the exact atomic state. In CCT the exact atomic 
state is defined as
\begin{equation}
|\Psi_i\rangle = e^{ T^{(0)}}|\Phi_i\rangle ,
\end{equation}
where $|\Phi_i\rangle$ is the reference state wave-function and 
$T^{(0)}$ is the unperturbed cluster operator, which incorporates the 
residual Coulomb interaction to all orders. We have introduced the superscript
to distinguish it from the second set of cluster operators, the perturbed
cluster operators, to be introduced later. In the case of a closed-shell 
ion, the model space of the ground state consists of a single Slater 
determinant, $|\Phi_0\rangle $, and $T^{(0)} = \sum_{i=1}^N T_{i}^{(0)}$, 
where $i$ is the order of excitation. However, in actual computations, 
incorporating $T_i^{(0)}$ with $i\geqslant 4$ is difficult with the existing 
computational resources. A simplified, but quite accurate approximation is 
the coupled-cluster single and double (CCSD) excitation approximation, in which 
\begin{equation}
  T^{(0)} = T_{1}^{(0)} + T_{2}^{(0)}.
\end{equation}
This is an approximation which embodies all the important electron correlation 
effects, and is a good starting point for structure and properties 
calculations of closed-shell ions. In the second quantized 
notations
\begin{subequations}
\begin{eqnarray}
T_1^{(0)} &= &\sum_{a,p} t_a^p {{a}_p^\dagger} a_a , \\
T_2^{(0)} &= &\frac{1}{2!}\sum_{a,b,p,q}t_{ab}^{pq} {{a}_p^\dagger}{{a}_q^
             \dagger}a_b a_a ,
\end{eqnarray}
\end{subequations}
where $t_{\ldots}^{\ldots}$ are cluster amplitudes, $a_i^{\dagger}$ ($a_i$)
are single particle creation (annihilation) operators and 
$abc\ldots$ ($pqr\ldots$) represent core (virtual) states. For the present
work, the ground state is the required atomic state 
$|\Psi_0\rangle = e^{{T^{(0)}}}|\Phi_0\rangle $ and satisfies the
eigenvalue equation 
\begin{equation}
H^{\rm DCB} e^{{T^{(0)}}}|\Phi_0\rangle = E_0e^{{T^{(0)}}}|\Phi_0\rangle ,
\end{equation}
where, $E_0$ and $|\Phi_0\rangle$ are the energy and reference state 
of the ground state, respectively. Following similar procedure, the CC 
eigenvalue equation of the one-valence \cite{mani-10}  and 
two-valence \cite{mani-11} systems may be defined.


\subsection{PRCC Theory}

 In the PRCC theory we introduce a new set of cluster operators,
$\mathbf{T^{(1)}}$, to incorporate an interaction Hamiltonian, $H_{\rm int}$, 
perturbatively. For general representation, we consider $\mathbf{T^{(1)}}$ as 
tensor operators of arbitrary rank and depends on the multipole 
structure of $H_{\rm int}$. The new cluster operators follow the selection 
rules associated with $H_{\rm int}$ and the modified ground state eigenvalue 
equation, after including the perturbation, is
\begin{equation}
(H^{\rm DCB}+\lambda H_{\rm int})|\tilde{\Psi}_0\rangle 
= \tilde{E}_0|\tilde{\Psi}_0\rangle , 
\end{equation}
where $\lambda$ is the perturbation parameter, $|\tilde{\Psi}_0\rangle $ is the
perturbed ground state and $\tilde{E}_0$ is the corresponding eigen energy. To 
calculate the electric dipole polarizability, $\alpha$, consider the 
perturbation as the interaction with an electrostatic field $\mathbf{E}$. The 
interaction Hamiltonian is then 
$H_{\rm int} =-\sum_i\mathbf{r}_i\cdot\mathbf{E} = \mathbf{D}\cdot\mathbf{E} $,
where $\mathbf{D}$ is the many electron electric dipole operator. Here,
$H_{\rm int}$ is odd in parity and to be more precise, $\mathbf{D}$, the 
operator in the electronic space is odd in parity and a rank one operator. 
Hence, the cluster opertors $\mathbf{T^{(1)}}$ are also rank one tensor 
operators and odd in parity, meaning, they connect states of different 
parities. Further more, the first energy correction 
$\langle \Psi_0|H_{\rm int}|\Psi_o\rangle = 0$ and therefore, 
$\tilde{E}_0=E_0$. We can then write, using PRCC theory, the perturbed
ground state as
\begin{eqnarray}
 |\tilde{\Psi}_0\rangle = e^{T^{(0)} + \lambda \mathbf{T}^{(1)}\cdot\mathbf{E}} 
 |\Phi_0\rangle = e^{T^{(0)}}\left [ 1 + \lambda \mathbf{T^{(1)}\cdot 
 \mathbf{E}} \right ] |\Phi_0\rangle ,
 \label{psi_tilde}
\end{eqnarray}
where, we have introduced the scalar product between $\mathbf{T}^{(1)}$ and
$\mathbf{E}$ for a consistent representation of the states and operators.
The advatage of introducing $\mathbf{T^{(1)}}$ and using 
$|\tilde{\Psi}_0\rangle $ is, it allows a systematic consolidation of
the correlation effects arising from multiple perturbations. 

  Based on the analysis of the low-order many-body perturbation theory 
diagrams, the single and double excitation operators of PRCC theory are 
represented as
\begin{subequations}
\begin{eqnarray}
  \mathbf{T}_1^{(1)} & = & \sum_{a,p} \tau_a^p \mathbf{C}_1 (\hat{r})
                       a_{p}^{\dagger}a_{a},
                            \\
  \mathbf{T}_2^{(1)} & = & \sum_{a,b,p,q} \sum_{l,k} \tau_{ab}^{pq}(l,k) 
                   \{ \mathbf{C}_l(\hat{r}_1) \mathbf{C}_k(\hat{r}_2)\}^{1}
                   a_{p}^{\dagger}a_{q}^{\dagger}a_{b}a_{a}. \;\;\;\;\;\;\;\;
\end{eqnarray}
\end{subequations}
where $\tau_{\ldots}^{\ldots}$ are the cluster amplitudes and 
$\mathbf{C}_i(\hat r)$ are $\mathbf{C}$-tensors of rank $i$. 
To represent $\mathbf{T}_1^{(1)}$, a rank one operator, we have used the 
$\mathbf{C}$-tensor of similar rank $\mathbf{C}_1(\hat r)$. And, the key 
difference of $\mathbf{T}_1^{(1)}$ from $T_1^{(0)}$ is $l_a+l_p$ must be odd, 
in other words $(-1)^{l_a+l_p} = -1$, where, $l_a$ ($l_p$) is the orbital 
angular momentum of the core (virtual) state $a$ ($p$). Coming to 
$\mathbf{T}_2^{(1)}$, to represent it two $\mathbf{C}$-tensor operators of 
rank $l$ and $k$ are coupled to a rank one tensor operator. In terms of
selection rules, the angular momenta of the orbitals and multipoles
in $\mathbf{T}_2^{(1)}$ must satisfy the triangular conditions 
$|j_a - j_p| \leqslant l \leqslant (j_a + j_p)$, 
$|j_b - j_q| \leqslant k \leqslant (j_b + j_q)$ and 
$|l - k| \leqslant 1 \leqslant (l + k)$. 
The other selection rule follows from the parity of $H_{\rm int}$, the
orbital angular momenta must satisfy the condition
$(-1)^{(l_a + l _p)} = - (-1)^{(l_b + l _q)}$.


\subsection{PRCC equations}
The ground state eigenvalue equation, in terms of the PRCC state, is 
\begin{equation}
  H^{\rm DCB} e^{\left [T^{(0)} + \lambda \mathbf{T}^{(1)}\cdot\mathbf{E}
  \right ]} |\Phi_0\rangle = E_0e^{\left [T^{(0)} 
  + \lambda \mathbf{T}^{(1)}\cdot\mathbf{E}\right ]} |\Phi_0\rangle .
 \label{prcc_eival}
\end{equation}
In the CCSD approximation we define the perturbed cluster operator 
$\mathbf{T}^{(1)}$ as
\begin{equation}
  \mathbf{T}^{(1)} = \mathbf{T}_1^{(1)} + \mathbf{T}_2^{(1)}.
  \label{prcc_ccsd}
\end{equation}
Using this, the PRCC equations are derived from Eq. (\ref{prcc_eival}). The 
derivation involves several operator contractions and these are more 
transparent with the normal ordered Hamiltonian
$H_{\rm N}^{\rm DCB} = H^{\rm DCB} - \langle \Phi_i|H^{\rm DCB}|\Phi_i\rangle$. 
The eigenvalue equation then assumes the form
\begin{equation}
  \left [H^{\rm DCB}_N  + \lambda H_{\rm int} \right ]|\tilde{\Psi}_0\rangle 
    = \left [E_0- \langle \Phi_0|H^{\rm DCB}|\Phi_0\rangle \right ]
    |\tilde{\Psi}_0\rangle .
\end{equation}
A more convenient form of the equation is 
\begin{equation}
  \left (H^{\rm DCB}_{\rm N} + \lambda H_{\rm int}\right )|\tilde{\Psi}_0\rangle
   = \Delta E_0|\tilde{\Psi}_0\rangle ,
\end{equation}
where, $\Delta E_0= E_0 - \langle \Phi_0|H^{\rm DCB}|\Phi_0\rangle$ is the 
ground state correlation energy. Following the definition in 
Eq. (\ref{psi_tilde}), the PRCC eigen-value equation is
\begin{equation}
  \left (H^{\rm DCB}_{\rm N} + \lambda H_{\rm int}\right )e^{T^{(0)}
    + \lambda \mathbf{T}^{(1)}\cdot\mathbf{E}} |\Phi_0\rangle =
    \Delta E_0e^{T^{(0)} + \lambda \mathbf{T}^{(1)}\cdot\mathbf{E}} 
    |\Phi_0\rangle .
\end{equation}
Applying $e^{-T^{(0)}}$ from the left, we get
\begin{equation}
  \left [\bar H^{\rm DCB}_{\rm N}  + \lambda \bar H_{\rm int}\right ] 
    e^{\lambda \mathbf{T}^{(1)\cdot\mathbf{E}}}|\Phi_0\rangle 
    = \Delta E_0e^{\lambda \mathbf{T}^{(1)}\cdot\mathbf{E}} |\Phi_0\rangle ,
  \label{prcc_eq1}
\end{equation}
where $\bar H^{\rm DCB}=e^{-T^{(0)}}H^{\rm DCB}e^{T^{(0)}}$ is the similarity 
transformed Hamiltonian. Multiplying Eq. (\ref{prcc_eq1}) from left by 
$e^{-\lambda \mathbf{T}^{(1)}}$ and considering the terms linear in $\lambda$, 
we get the PRCC equation
\begin{equation}
   \left [\bar{H}^{\rm DC}_{\rm N},\mathbf{T}^{(1)}\right ]\cdot\mathbf{E} 
    + \bar{H}_{\rm int}|\Phi_0\rangle = 0 .
\end{equation}
Here, the similarity transformed interaction Hamiltonian $\bar{H}_{\rm int}$ 
terminates at second order as $H_{\rm int} $ is a one-body interaction 
Hamiltonian. Expanding $\bar H_{\rm int}$ and dropping $\mathbf{E}$ for 
simplicity, the  PRCC equation assumes the form
\begin{eqnarray}
  \left [\bar{H}_{\rm N}^{\rm DCB},\mathbf{T}^{(1)}\right ] |\Phi_0\rangle
     & = & \bigg ( \mathbf{D} + \left [\mathbf{D},T^{(0)}\right ]  
     \nonumber \\
  && + \frac{1}{2}\left[ \left[\mathbf{D},T^{(0)}\right ], T^{(0)}
     \right ] \bigg )|\Phi_0\rangle . \;\;\;
\end{eqnarray}
The equations of $\mathbf{T}_1^{(1)}$ are obtained after projecting
the equation on singly excited states $\langle \Phi_a^p|$. These excitation
states, however, must be opposite in parity to $|\Phi_0 \rangle $.  
The $\mathbf{T}_2^{(1)}$ equations are obtain in a similar way after 
projecting on the doubly excited states $\langle \Phi_{ab}^{pq}|$. The 
equations form a set of coupled nonlinear algebraic equations. The equations
and a description of the different terms along with diagrammatic analysis
are given in our previous works \cite{chattopadhyay-12a,chattopadhyay-12b}. An 
approximation which incorporates all the important many-body effects like 
core-polarization is the linearized PRCC (LPRCC). In this approximation, only 
the terms linear in $T^{(0)}$, equivalent to retaining only 
$\{\contraction[0.5ex]{}{H}{_{\rm N}}{T}H_{\rm N}T^{(1)}\}$ and 
$\{\contraction[0.5ex]{}{\mathbf{r}}{_i}{T}\mathbf{r}_iT^{(0)} \}$ in 
the PRCC equations. 
\begin{figure}[h]
 \includegraphics[width=8cm]{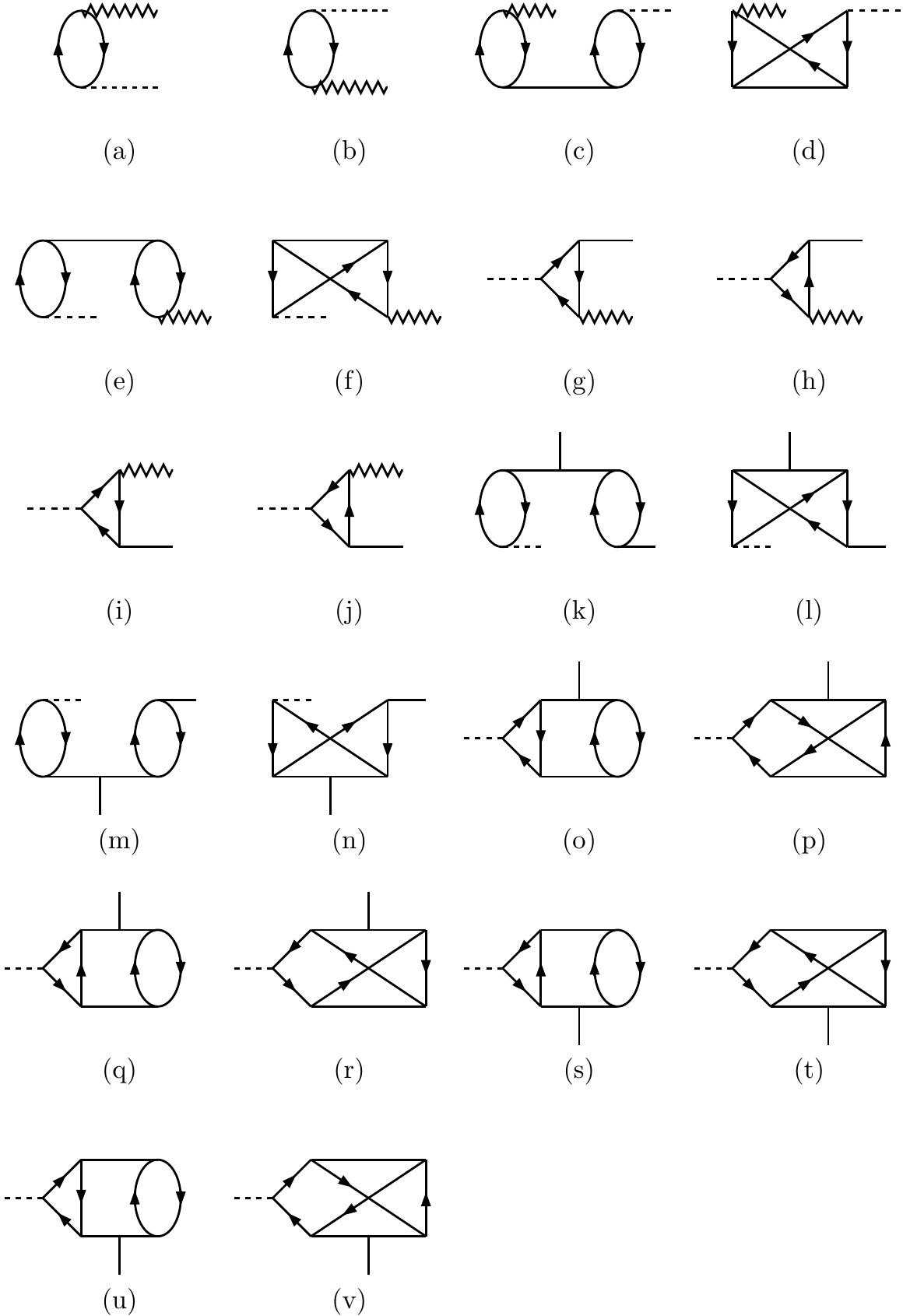}
 \caption{Diagrams of the $\alpha$ in the PRCC theory. The single excitation
          operators with a wavy line represent $\mathbf{T}^{(1)}_1$. 
          Similarly, the double excitation diagrams with an extra vertical 
          line represent $\mathbf{T}^{(1)}_2$.
         }
 \label{dip_pol_diag}
\end{figure}


\section{Dipole Polarizability}

In the PRCC theory we can write the $\alpha$ of the ground state of a
closed-shell atom as \cite{chattopadhyay-12a,chattopadhyay-12b}
\begin{equation}
  \alpha = -\frac{\langle \Phi_0|\mathbf{T}^{(1)\dagger}\bar{\mathbf{D}} + 
   \bar{\mathbf{D}}\mathbf{T}^{(1)}|\Phi_0\rangle}{\langle\Psi_0|\Psi_0\rangle},
\end{equation}
where, $\bar{\mathbf{D}} = e^{{T}^{(0)\dagger}}\mathbf{D} e^{T^{(0)}}$, 
represents the unitary transformed electric dipole operator. Retaining only
the dominant terms, we obtain
\begin{eqnarray}
 \alpha &=& -\frac{1}{\cal N}\langle\Phi_0|\mathbf{T}_1^{(1)\dagger}\mathbf{D} 
       + \mathbf{D}\mathbf{T}_1^{(1)} 
       + \mathbf{T}_1^{(1)\dagger}\mathbf{D}T_2^{(0)}
       + T_2^{(0)\dagger}\mathbf{D}\mathbf{T}_1^{(1)} \nonumber \\
    && + \mathbf{T}_1^{(1)\dagger}\mathbf{D}T_1^{(0)} 
       + T_1^{(0)\dagger}\mathbf{D}\mathbf{T}_1^{(1)}
       + \mathbf{T}_2^{(1)\dagger}\mathbf{D}T_1^{(0)} \nonumber \\
    && + T_1^{(0)\dagger}\mathbf{D}\mathbf{T}_2^{(1)}
       + \mathbf{T}_2^{(1)\dagger}\mathbf{D}T_2^{(0)} 
       + T_2^{(0)\dagger}\mathbf{D}\mathbf{T}_2^{(1)}
     |\Phi_0\rangle, 
  \label{dip_pol_eqn}
\end{eqnarray}
where ${\cal N} = \langle\Phi_0|\exp[T^{(0)\dagger}]\exp[T^{(0)}]
|\Phi_0\rangle$ is the normalization factor, which involves a non-terminating
series of contractions between ${T^{(0)}}^\dagger $ and $T^{(0)} $. However, in 
the present work we use 
${\cal N} \approx \langle\Phi_0|1 + T_1^{(0)\dagger}T_1^{(0)} + 
T_2^{(0)\dagger}T_2^{(0)}|\Phi_0\rangle$. An evident advantage of computing 
$\alpha$ with PRCC theory is the absence of summation over the intermediate 
states $|\Psi_I\rangle $. The summation is subsumed in the evaluation of 
$\mathbf{T}^{(1)}$ in a natural way and eliminates the need for a complete 
set of intermediate states. 

  For further analysis and evaluation of the different terms in 
Eq. (\ref{dip_pol_eqn}), we use many-body diagrams or Goldstone diagrams. To 
evaluate the diagrams we follow the notations and conventions given 
in ref. \cite{lindgren-86}. However, there is an additional feature in the 
diagrams of  $\alpha$, we employ a wavy interaction line to represent the 
diagrams of $\mathbf{T}^{(1)}_1 $, so that it is different from the diagrams 
of $T_1^{(0)}$. Similarly, to represent $\mathbf{T}^{(1)}_2 $ we introduce 
a vertical line to the interaction line. After due consideration of the 
equivalent diagrams, the terms in  Eq. (\ref{dip_pol_eqn}) correspond to 22 
unique Goldstone diagrams and these are shown in Fig. \ref{dip_pol_diag}. 
The equivalent algebraic expression is 
\begin{eqnarray}
  \alpha & = & \sum_{ap} \left ({\tau_a^p}^* d_{ap} + d_{ap}^*\tau_a^p 
     \right ) + \sum_{abpq}\left [ ( {\tau_a^p}^*d_{bq}^* 
     + {\tau_a^q}^*d_{bp}^* ) t_{ab}^{pq}  \right . 
                   \nonumber \\
   & & \left . + \tilde{t}_{ab}^{pq*}d_{ap}\tau_b^q \right ]
     + \sum_{apq}({t_a^q}^*d_{pq}\tau_a^p  + {\tau_a^q}^*d_{pq}t_a^p )
                    \nonumber \\
   & & - \sum_{abp}({t_b^p}^*d_{ab}\tau_a^p  + {\tau_b^p}^*d_{ab}t_a^p )
     + \sum_{abpq} (\tilde{\tau}_{ab}^{pq*}d_{bq}t_b^q 
                    \nonumber \\
   & & + d_{bq}^*t_b^{q*} \tilde{\tau}_{ab}^{pq} )
     + \sum_{abpqr}(\tilde{\tau}_{ab}^{rq*}d_{pr}t_{ab}^{pq}
     + \tilde{t}_{ab}^{rq*}d_{pr}\tau_{ab}^{pq} )
                    \nonumber \\
   & & + \sum_{abcpq}(\tilde{\tau}_{cb}^{pq*}d_{ca}t_{ab}^{pq}
     + \tilde{t}_{cb}^{pq*}d_{ca}\tau_{ab}^{pq} ),
\end{eqnarray}
where $d_{ab} = \langle a|d|b\rangle$, and 
$\tilde{\tau}_{ab}^{pq} = \tau_{ab}^{pq} - \tau_{ab}^{qp} $  and
$\tilde{t}_{ab}^{pq} = t_{ab}^{pq} - t_{ab}^{qp} $  are the antysymmetrized
cluster amplitudes.

In the figure, the first two diagrams, Fig. \ref{dip_pol_diag}(a) and 
\ref{dip_pol_diag}(b), are the most important ones. These represent 
$\mathbf{T}_1^{(1)\dagger}\mathbf{D}$ and  $\mathbf{D}\mathbf{T}_1^{(1)}$, 
respectively, and subsume DF and the effects of random phase approximation 
(RPA). The next two diagrams in the figure, Fig.\ref{dip_pol_diag}(c) and 
Fig.\ref{dip_pol_diag}(d), arise from the term 
$\mathbf{T}_1^{(1)\dagger}\mathbf{D}T_2^{(0)}$. Similarly, the diagrams
in Fig.\ref{dip_pol_diag}(e-f) correspond to the 
hermitian conjugate, $T_2^{(0)\dagger}\mathbf{D}\mathbf{T}_1^{(1)}$. 
These are the two leading order terms among the second order contributions, in 
terms of the cluster amplitudes, to $\alpha$. The reason is, both
the terms consist of dominant RCC and PRCC amplitudes, 
$T_2^{(0)} $ and $\mathbf{T}_1^{(1)} $, respectively. 

 Among the second order contributions, the next to leading order terms are
$\mathbf{T}_2^{(1)\dagger}\mathbf{D}T_2^{(0)}$ and 
$T_2^{(0)\dagger}\mathbf{D}\mathbf{T}_2^{(1)}$. Each of these terms generate
four diagrams, which are given in Fig.\ref{dip_pol_diag}(o-r) and 
Fig.\ref{dip_pol_diag}(s-v) and these correspond to 
$\mathbf{T}_2^{(1)\dagger}\mathbf{D}T_2^{(0)}$ and 
$T_2^{(0)\dagger}\mathbf{D}\mathbf{T}_2^{(1)}$, respectively. The remaining
second order terms, $\mathbf{T}_1^{(1)\dagger}\mathbf{D}T_1^{(0)}$,
$\mathbf{T}_2^{(1)\dagger}\mathbf{D}T_1^{(0)}$ and their hermitian
conjugates, have marginal contributions to $\alpha$. However, for 
completeness, these are included in the computations.
\begin{table}[h]
   \caption{The $\alpha_0$ and $\beta$ parameters of the even tempered
            GTO basis used in the present calculations.}
   \label{basis}
   \begin{tabular}{cccccccc}
   \hline
   \hline
     Atom & \multicolumn{2}{c}{$s$} & \multicolumn{2}{c}{$p$} &
     \multicolumn{2}{c}{$d$}  \\
     & $\alpha_{0}$  & $\beta$ & $\alpha_{0}$ & $\beta$
     & $\alpha_{0}$  & $\beta$  \\
     \hline
     $\rm{Na}^{+}$ &\, 0.0025  &\, 2.210 &\, 0.00955 &\, 2.125 &\, 0.00700
     &\, 2.750 \\
     $\rm{K}^{+}$  &\, 0.0055  &\, 2.250 &\, 0.00995 &\, 2.155 &\, 0.00690
     &\, 2.550 \\
     $\rm{Rb}^{+}$ &\, 0.0052  &\, 2.300 &\, 0.00855 &\, 2.205 &\, 0.00750
     &\, 2.145 \\
     $\rm{Cs}^{+}$ &\, 0.0097  &\, 2.050 &\, 0.00975 &\, 2.005 &\, 0.00995
     &\, 1.705 \\
     $\rm{Fr}^{+}$ &\, 0.0068  &\, 2.110 &\, 0.00645 &\, 2.050 &\, 0.00985
     &\, 1.915 \\
     \hline
   \end{tabular}
\end{table}


\section{Calculational details}

\subsection{Basis set and nuclear density}
 The first step of our computations, which is also true of any atomic and 
molecular computations, is to generate an spin-orbital basis set. For the 
present work, the basis set is even-tempered gaussian type orbitals (GTOs) 
\cite{mohanty-90} generated with the Dirac-Hartree-Fock Hamiltonian. This
means, the radial part of the spin-orbitals are linear combinations of the 
Gaussian type functions. The Gaussian type functions which constitutes the
large components are of the form
\begin{equation}
   g_{\kappa p}^{L}(r) = C^{L}_{\kappa i} r^{n_{\kappa}}e^{-\alpha_{p}r^{2}},
\end{equation}
where $p=0,1\ldots m$ is the GTO index and $m$ is the number of gaussian type 
functions. The exponent $\alpha_{p} = \alpha_{0} \beta^{p-1}$, where 
$\alpha_{0}$ and $\beta$ are two independent paramters. Similarly, the small 
components of the spin-orbitals are linear combination of 
$g_{\kappa p}^{S}(r)$, which are generated from $g_{\kappa p}^{L}(r)$ through 
the kinetic balance condition \cite{stanton-84}. The GTOs are calculated
on a grid \cite{chaudhuri-99} with optimized values of $\alpha_{0}$ and 
$\beta$. The optimization is done for individual atoms to match the 
spin-orbital energies and self consistent field (SCF) energy of GRASP92 
\cite{parpia-96}. For the current work, the optimized $\alpha_{0}$ and 
$\beta$ are listed in Table. \ref{basis}. For comparison, the spin-orbital
energies of Cs$^+$ obtained from the GTO and GRASP92 are listed in 
Table \ref{orb_cs}. In the table, the deviation of the GTO results from the 
GRASP92 is $\sim 10^{-3}$, which is quite small. We obtain similar level
of deviations for the other ions as well.

%

\begin{table}[h]
   \caption{Core orbital energies of $\rm{Cs}^{+}$ in atomic units.}
   \label{orb_cs}
   \begin{tabular}{ldd}
   \hline
     Orbital & \multicolumn{1}{c}{DC}  &
               \multicolumn{1}{c}{GRASP92 \cite{parpia-96}}   \\ \hline
     $1s_{1/2}$   & -1330.1173  &  -1330.1129    \\
     $2s_{1/2}$   & -212.5643   &  -212.5673   \\
     $2p_{1/2}$   & -199.4294   &  -199.4288    \\
     $2p_{3/2}$   & -186.4366   &  -186.4358    \\
     $3s_{1/2}$   & -45.9697    &  -45.9695    \\
     $3p_{1/2}$   & -40.4483    &  -40.4455     \\
     $3p_{3/2}$   & -37.8943    &  -37.8917     \\
     $3d_{3/2}$   & -28.3096    &  -28.3030     \\
     $3d_{5/2}$   & -27.7752    &  -27.7682     \\
     $4s_{1/2}$   & -9.5128     &  -9.5106     \\
     $4p_{1/2}$   & -7.4463     &  -7.4437      \\
     $4p_{3/2}$   & -6.9209     &  -6.9188      \\
     $4d_{3/2}$   & -3.4856     &  -3.4921      \\
     $4d_{5/2}$   & -3.3969     &  -3.4038      \\
     $5s_{1/2}$   & -1.4898     &  -1.4933      \\
     $5p_{1/2}$   & -0.9079     &  -0.9139      \\
     $5p_{3/2}$   & -0.8403     &  -0.8459      \\
   \hline
   \end{tabular}
\end{table}

The next step, related to spin-orbial basis set, is the choice of an ideal
basis set size. For this, we examine the convergence of $\alpha $ using
the LPRCC theory. We calculate $\alpha$ starting with a basis set of 50 GTOs 
and increase the basis set size in steps through a series of calculations. 
The results of the such a series of calculations are listed in 
Table. \ref{basis_cs}, it shows the convergence of $\alpha$ of Cs$^+$ as a 
function of basis set size.  

In the present work we have considered finite size Fermi density distribution 
of the nucleus
\begin{equation}
   \rho_{\rm nuc}(r) = \frac{\rho_0}{1 + e^{(r-c)/a}},
\end{equation}
where, $a=t 4\ln(3)$. The parameter $c$ is the half charge radius so that 
$\rho_{\rm nuc}(c) = {\rho_0}/{2}$ and $t$ is the skin thickness. Coming to the
PRCC equations, these are solved iteratively using Jacobi method, we have 
chosen this method as it is parallelizable. The method, however, is slow to 
converge. So, to accelerate the convergence we use direct inversion in the 
iterated subspace (DIIS)\cite{pulay-80}. 

\begin{table}[h]
  \caption{Convergence pattern of $\alpha$ ($\rm{Cs}^{+}$) as a function of
           the basis set size.}
  \label{basis_cs}
  \begin{tabular}{lcc}
      \hline
      No. of orbitals & Basis size & $\alpha $   \\
      \hline
      103  & $(15s, 13p, 13d,   9f,   9g)  $ & 14.9480  \\
      117  & $(17s, 15p, 15d,  11f,   9g)  $ & 14.9235  \\
      131  & $(19s, 17p, 17d,  11f,  11g)  $ & 14.9124  \\
      143  & $(23s, 19p, 19d,  13f,  11g)  $ & 14.9086  \\
      159  & $(23s, 21p, 21d,  13f,  13g)  $ & 14.9077  \\
      177  & $(25s, 23p, 23d,  15f,  15g)  $ & 14.9077  \\
      \hline
  \end{tabular}
\end{table}


\subsection{Intermediate Diagrams}

   The PRCC diagrams corresponding to the nonlinear terms
are numerous and topologically complex. Further more, in these diagrams,
the number of the spin-orbitals involved is large and in general, the diagrams 
with the largest number of spin-orbitals are associated with the terms 
$H_NT_2^{(0)}T_2^{(0)}$, $H_NT_1^{(0)}T_1^{(0)}T_2^{(0)}$ 
and $H_NT_1^{(0)}T_1^{(0)}T_1^{(0)}T_1^{(0)}$. All of these terms have a 
common feature: the presence of the Coulomb integral 
$\langle ab|1/r_{12}|pq\rangle$. Returning to the number of spin-orbitals, 
the $T_2^{(0)}$ diagrams arising from any of the three terms 
mentioned earlier consist of four core and virtual spin-orbitals each. 
Accordingly, the number of times a diagram is evaluated, $N_d$, scales as 
$n_o^4n_v^4$ and sets the computational requirements. Here, $n_o$ and $n_v$ 
are the number of core and virtual spin-orbitals, respectively. In the present 
work, $n_o\sim 10$ and $n_v\sim 100$ for lighter atoms and moderate sized 
basis sets, even then $N_d\sim 10^{12}$. This is a large number and puts a 
huge constraint on the computational resources. 
\begin{figure}[h]
 \includegraphics[width=8cm]{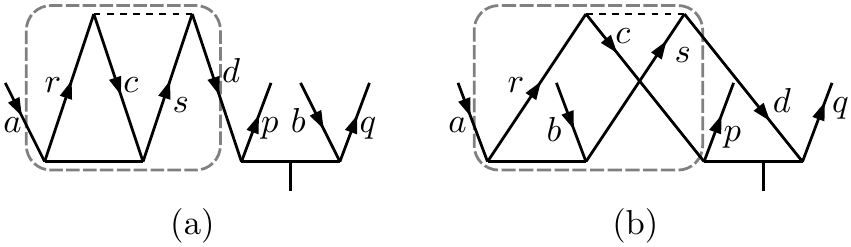}
 \caption{Example diagrams of $H_NT_2^{(0)}T_2^{(0)}$ which contribute to the
          $\mathbf{T}_2^{(1)} $ equations. The portion of the diagrams within
          the rectangles with rounded corners are examples of the one body 
          (a) and two-body (b) intermediate diagrams.
         }
 \label{cc_ims}
\end{figure}

 To mitigate the computational constraints arising from the $n_o^4n_v^4$
scaling, we separate the diagrams into two parts. One of the parts scales
at the most $n_o^2n_v^4$ and the total diagram is equivalent to the product 
of the parts. The part of the diagram, which is calculated first is referred 
to as the intermediate diagram. During computations, all the intermediate 
diagrams are calculated first and stored. Later, these are used to combine 
with the remaining part of the RCC diagram and the total diagram is 
calculated. The scaling is still $n_o^2n_v^4$ and compared to the 
$n_o^4n_v^4$ scalling, this improves the performance by several orders of 
magnitudes. 

To examine in more detail, consider the term $H_NT_2^{(0)}\mathbf{T}_2^{(1)}$, 
the algebraic expression for one of the terms contributing to the 
$\mathbf{T}_2^{(0)}$ is
\begin{equation}
  \left ( \tau_{ab}^{pq}\right )_{\rm 2a} a_p^\dagger a_q^\dagger a_ba_a   = 
        \sum_{rcsd}t_{ac}^{rs} v_{rs}^{cd}\tau_{db}^{pq} 
        a_p^\dagger a_q^\dagger a_ba_a ,
  \label{ims_sing}
\end{equation} 
and it is diagrammatically equivalent to Fig. \ref{cc_ims}(a). However, while 
evaluating the diagram, the part within the dashed round rectangle or the 
intermediate diagram can be separated and computed first. The 
Eq. (\ref{ims_sing}) can then be written as 
\begin{equation}
  \left ( \tau_{ab}^{pq}\right )_{\rm 2a}a_p^\dagger a_q^\dagger a_ba_a  = 
          \sum_d \left ( \eta_a^d a_d^\dagger a_a \right ) \Big ( 
          \tau_{db}^{pq} a_p^\dagger a_q^\dagger a_ba_d \Big ),
\end{equation}
where $\eta_a^d=\sum_{rcs}t_{ac}^{rs} v_{rs}^{cd}$ is the amplitude of the 
effective one-body operator corresponding to the intermediate diagram. It 
scales as $n_o^3n_v^2$ and when contracted with $\mathbf{T}_2^{(1)}$, the 
computation still cales as $n_o^3n_v^2 $. This is much less than the 
$n_o^4n_v^4$ scaling. Consider another term
\begin{equation}
   \left ( \tau_{ab}^{pq}\right )_{\rm 2b} a_p^\dagger a_q^\dagger a_ba_a   = 
        \sum_{rcsd}t_{ab}^{rs} v_{rs}^{cd}\tau_{cd}^{pq} 
        a_p^\dagger a_q^\dagger a_ba_a ,
  \label{ims_dbl}
\end{equation}
and it is diagrammatically equivalent to Fig. \ref{cc_ims}(b). Like in the 
previous case, the intermediate diagram ( part within the dashed 
round-rectangle ) can be calculated first and the equation can rewritten as
\begin{equation}
   \left ( \tau_{ab}^{pq}\right )_{\rm 2b} a_p^\dagger a_q^\dagger a_ba_a   = 
      \sum_{cd}\left (\eta_{ab}^{cd} a_c^\dagger a_d^\dagger a_ba_a \right ) 
      \Big ( \tau_{cd}^{pq} a_p^\dagger a_q^\dagger a_da_c\Big ).
\end{equation}
Here, the intermediate diagram corresponds to a two-body effective operator
with amplitude $\eta_{ab}^{cd} =\sum_{rs}t_{ab}^{rs} v_{rs}^{cd}$ and scales
as $n_o^4n_v^2$. The scaling remains the same when the total diagram is 
evaluated. Extending the method to other diagrams, depending on the 
topology, there are other forms of one-body and two-body intermediate diagrams.
\begin{figure}[h]
 \includegraphics[width=8cm]{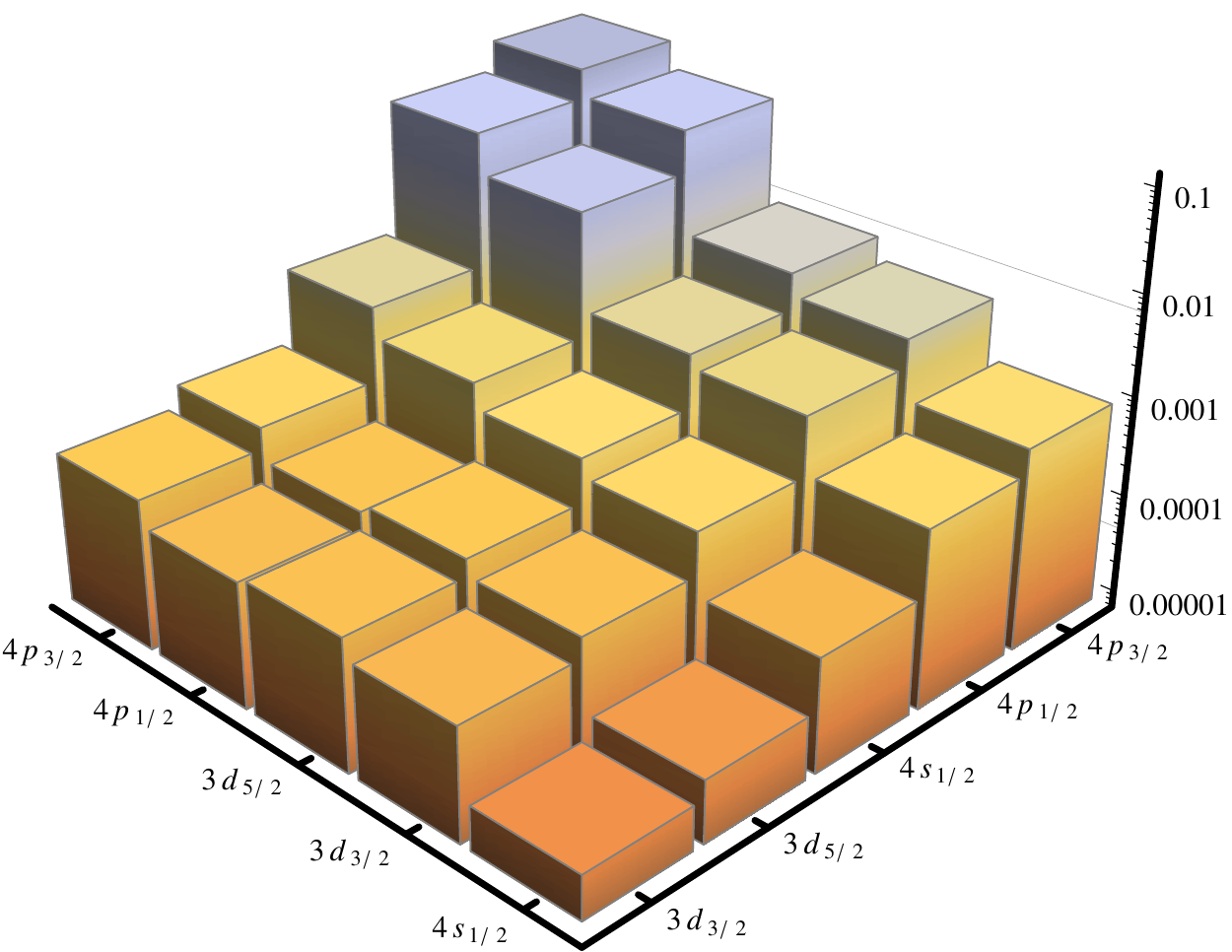}
 \label{rb_chart}
 \caption{Contributions to the next to leading order terms 
          $\mathbf{T}_1{^{(1)\dagger}} \mathbf{D}T_2^{(0)}$  + h.c. in terms
          of the pairs of core spin-orbitals. }
\end{figure}


\section{Results and Discussions}

  To compute $\alpha$ using PRCC theory, as described earlier, we consider
terms up to second order in the cluster operators. We have, however, studied 
terms which are third order in cluster operators and examined the contributions 
from the leading order terms. But the contributions are negligible and this
validates our choice of considering terms only upto second order in cluster
operators. To begin with, we compute $\alpha$ using the cluster amplitude 
obtained from the LPRCC and results are presented in 
Table \ref{linearised_polarizabilty}. In the table we have listed, for 
systematic comparison, the experimental data and results from previous 
theoretical computations. 
\begin{table}[h]
        \caption{Static dipole polarizabilty of alkali ions}
        \label{linearised_polarizabilty}
        \begin{center}
        \begin{tabular}{lccccc}
            \hline
            Atom & LPRCC & RCCSDT\cite{lim-02} & RRPA \cite{johnson-83}
            &  Expt. \\ \hline
            $\rm{Na}^{+}$ & 1.009  & 1.00(4) & 0.9457 &  0.9980(33) 
            \cite{gray-88}      \\
            $\rm{K}^{+}$  & 5.521  & 5.52(4) & 5.457  &  5.47(5)
            \cite{opik-67}      \\
            $\rm{Rb}^{+}$ & 8.986  & 9.11(4) & 9.076  &  9.0
            \cite{johansson-61} \\
            $\rm{Cs}^{+}$ & 14.924 & 15.8(1) & 15.81  &  15.644(5)  
            \cite{zhou-89}      \\
            $\rm{Fr}^{+}$ & 19.506 & 20.4(2) &        &             \\
            \hline
        \end{tabular}
        \end{center}
\end{table}

For $\rm{Na}^{+}$ and $\rm{K}^{+}$, our values of $\alpha$  are higher than
the experimental values by 1\% and 0.9\% respectively. However, for 
$\rm{Rb}^{+}$ and $\rm{Cs}^{+}$ our results are lower than the experimental 
values by 0.15\% and 4.8\%, respectively. In terms of theoretical results,
our results of $\rm{Na}^{+}$ and $\rm{K}^{+}$ are in excellent agreement with
the previous work which used the RCCSDT method for computation. But, for
$\rm{Rb}^{+}$ and $\rm{Cs}^{+}$, like in the experimental data, our results 
are lower than the RCCSDT results. One possible reason for these deviations 
in the heavier ions could be the exclusion of triple excitation cluster 
operators in the present work. Our result of Fr$^+$ seems to bear out this 
reasoning as the same trend is observed ( our result is 4.4\% lower than the 
RCSSDT result)  in this case as well. However, in absence of experimental 
data for Fr$^+$, it is difficult to arrive at a definite conclusion.

   To investigate the importance of Breit interaction, $H_{\rm B}$, in 
computing $\alpha$ of the alkali ions, we exclude $H_{\rm B}$ in the atomic  
Hamiltonian and do a set of systematic calculations. Our results for the
values of $\alpha$ are then 1.008, 5.514, 8.973 and 14.908 for $\rm{Na}^{+}$, 
$\rm{K}^{+}$, $\rm{Rb}^{+}$ and $\rm{Cs}^{+}$, respectively. These values are 
0.001, 0.007, 0.013 and 0.016 a.u lower than the results computed using
the Dirac-Coulomb-Breit Hamiltonian. This indicates that the correction from
the Breit interaction is larger in heavier ions and this is as expected since 
the stronger nuclear potential in heavier ions translates to larger Breit
correction. 

   For a more detailed study, we examine the contributions from  each of
the terms in the Eq. \ref{dip_pol_eqn} and these are listed in 
Table.  \ref{result_lprcc}. 
\begin{table}[h]
    \caption{Contribution to $\alpha $ from different terms and their
             hermitian conjugates in the linearized PRCC theory.}
    \label{result_lprcc}
    \begin{center}
    \begin{tabular}{lddddd}
        \hline
        Terms + h.c. & \multicolumn{1}{r}{$\rm{Na}^{+}$}
        & \multicolumn{1}{r}{$\rm{K}^{+}$}
        & \multicolumn{1}{r}{$\rm{Rb}^{+}$}
        & \multicolumn{1}{r}{$\rm{Cs}^{+}$}
        & \multicolumn{1}{r}{$\rm{Fr}^{+}$}  \\
        \hline
        $\mathbf{T}_1^{(1)\dagger}\mathbf{D} $ 
        & 1.018 & 6.043 & 10.029 & 17.472 & 22.926 \\
        $\mathbf{T}_1{^{(1)\dagger}} \mathbf{D}T_2^{(0)} $  
        & -0.018 & -0.299 & -0.519 & -1.023 &  -1.326 \\
        $\mathbf{T}_1{^{(1)\dagger}}\mathbf{D}T_1^{(0)} $ 
        & 0.012 & -0.038 & -0.072 & -0.188 &  -0.126  \\
        $\mathbf{T}_2{^{(1)\dagger}}\mathbf{D}T_1^{(0)} $ 
        & -0.001 & 0.008 & 0.016 & 0.039 &  0.026 \\
        $\mathbf{T}_2{^{(1)\dagger}}\mathbf{D}T_2^{(0)} $  
        & 0.023  & 0.204 & 0.332  & 0.654 & 0.834 \\
        Normalization & 1.025 & 1.072 & 1.089 & 1.136 & 1.145  \\
        Total & 1.009 & 5.521  & 8.986 & 14.924 & 19.506 \\
        \hline
    \end{tabular}
    \end{center}
\end{table}
The leading order contribution arises from 
$\mathbf{T}_1^{(1)\dagger}\mathbf{D} $ + h.c  and diagrammaticaly, it 
corresponds to the first two diagrams in Fig. \ref{dip_pol_diag} . These are
also the lowest order terms and are the dominant terms since these subsume the 
contributions from the Dirac-Fock and RPA effects. For all the 
ions, the results from the dominant terms exceeds the final results. Here, 
it must be mentioned that a similar trend is observed in the results of
noble gas atoms as well \cite{chattopadhyay-12a,chattopadhyay-12b}.  The next 
to leading order (NLO) contributions arise from 
the $\mathbf{T}_1{^{(1)\dagger}} \mathbf{D}T_2^{(0)} $ + h.c. The 
contributions from these terms are an order of magnitude smaller then
$\mathbf{T}_1^{(1)\dagger}\mathbf{D} $ + h.c but more importantly,
the contributions are opposite in phase. Interestingly, the next important 
terms $\mathbf{T}_2{^{(1)\dagger}} \mathbf{D}T_2^{(0)} $ + h.c have
contributions which nearly cancels the NLO contributions. Continuing further,
among the second order terms, the smallest contribution arise from 
$\mathbf{T}_2{^{(1)\dagger}}\mathbf{D}T_1^{(0)} $ + h.c., which is perhaps not
surprising since $\mathbf{T}_2{^{(1)\dagger}}$ and $T_1^{(0)}$ are the 
cluster operators with smaller amplitudes in PRCC and RCC theories, 
respectively. Collecting the results, the net contributions from the
second order terms are 0.016, -0.117, -0.223, -0.456 and -0.517 for 
$\rm{Na}^{+}$, $\rm{K}^{+}$, $\rm{Rb}^{+}$, $\rm{Cs}^{+}$ and $\rm{Fr}^{+}$, 
respectively. 

   Next, we consider all the terms in the PRCC theory, including the terms
which are non-linear in cluster operators. The results of $\alpha$
are presented in the Table \ref{result_prcc}.
\begin{table}[h]
    \caption{Contribution to $\alpha $ from different terms and their
             conjugate in the  PRCC theory after including the terms
             nonlinear in cluster operators.}
    \label{result_prcc}
    \begin{center}
    \begin{tabular}{ldddd}
        \hline
        Terms + h.c.  & \multicolumn{1}{r}{$\rm{Na}^{+}$}
        & \multicolumn{1}{r}{$\rm{K}^{+}$}
        & \multicolumn{1}{r}{$\rm{Rb}^{+}$}
        & \multicolumn{1}{r}{$\rm{Cs}^{+}$} \\
        \hline
        $\mathbf{T}_1^{(1)\dagger}\mathbf{D} $ 
        & 1.034 & 6.302 & 10.438 & 18.376   \\
        $\mathbf{T}_1{^{(1)\dagger}} \mathbf{D}T_2^{(0)} $ 
        & -0.018 & -0.316 & -0.544 & -1.084 \\
        $\mathbf{T}_1{^{(1)\dagger}}\mathbf{D}T_1^{(0)} $
        & 0.012 & -0.040 & -0.076 & -0.198  \\
        $\mathbf{T}_2{^{(1)\dagger}}\mathbf{D}T_1^{(0)} $
        & -0.0008 & 0.008 & 0.016 & 0.038   \\
        $\mathbf{T}_2{^{(1)\dagger}}\mathbf{D}T_2^{(0)} $
        & 0.024  & 0.194 & 0.308  & 0.596   \\
        Normalization & 1.026 & 1.072 & 1.090 & 1.136 \\
        Total & 1.025 & 5.735  & 9.305 & 15.606 \\
        \hline
    \end{tabular}
    \end{center}
\end{table}
For $\rm{Na}^{+}$ the result of $\alpha$ is 2.6\% higher than the
experimental value. Similarly,  for $\rm{K}^{+}$ and $\rm{Rb}^{+}$ the results 
are 4.6\% and 3.3\% higher than the experimental values. For
$\rm{Cs}^{+}$ the nonlinear PRCC theory gives a much improved result than
the LPRCC results and the deviation from the experimental value is
reduced to 0.24\%. On a closer examination, most of the change associated
with the nonlinear PRCC can be attributed to the increased contribution from
$\mathbf{T}_1^{(1)\dagger}\mathbf{D} $ + h.c. As these terms subsume RPA
effects, the increased contributions indicate that RPA effects are larger in 
the nonlinear PRCC. 

To investigate the RPA effects in detail, we isolate the contributions
from each of the core spin-orbitals to 
$\mathbf{T}_1^{(1)\dagger}\mathbf{D} $ + h.c.
and The dominant contributions are presented in Table. \ref{result_t1d}.
\begin{table}[h]
    \caption{Four leading contributions to 
        $\{ \mathbf{T}_1^{(1)\dagger}\mathbf{D} + \rm{h.c} \}$ to $\alpha $
        in terms of the core spin-orbitals. }
    \label{result_t1d}
    \begin{center}
    \begin{tabular}{rrr}
        \hline
          \multicolumn{1}{c}{$\rm{Na}^{+}$} & \multicolumn{1}{c}{$\rm{K}^{+}$}
        & \multicolumn{1}{c}{$\rm{Rb}^{+}$}  \\ \hline
        0.652 (2$p_{3/2}$) & 4.016 (3$p_{3/2}$) & 6.858 (4$p_{3/2}$) \\
        0.322 (2$p_{1/2}$) & 1.938 (3$p_{1/2}$) & 3.038  (4$p_{1/2}$) \\
        0.044 (2$s_{1/2}$) & 0.076 (3$s_{1/2}$) & 0.058 (4$s_{1/2}$) \\
        0.0004 (1$s_{1/2}$) & 0.008 (2$p_{3/2}$) & 0.044 (3$d_{5/2}$) \\
        \hline
       \multicolumn{1}{c}{$\rm{Cs}^{+}$}   & \multicolumn{1}{c}{$\rm{Fr}^{+}$}
      &  \\  \hline
          12.375 (5$p_{3/2}$) & 18.287 (6$p_{3/2}$) & \\
          4.735 (5$p_{1/2}$) & 4.073 (6$p_{1/2}$) & \\
          0.192 (4$d_{5/2}$) & 0.376 (5$d_{5/2}$) & \\
         0.121 (4$d_{3/2}$) & 0.211 (5$d_{3/2}$) & \\ \hline
    \end{tabular}
    \end{center}
\end{table}
It is to be noted that $\alpha$ has a quadratic dependence on the radial 
distance, so the orbitals with larger spatial extension contribute dominantly. 
The effect of this is discernible in the results, for all the alkali ions the 
leading contribution to $\alpha$ comes from the outermost $np_{3/2}$ orbital,
which is the occupied orbital with largest radial extent. The next leading
contribution arise from the $np_{1/2}$ orbital. An important observation is,
as we proceed from from lower Z to higher Z, the ratio of contribution of 
$np_{3/2}$ to the $np_{1/2}$ increases. It is 1.8, 2.1, 2.3, 2.6 and 4.5 for 
$\rm{Na}^{+}$, $\rm{K}^{+}$, $\rm{Rb}^{+}$, $\rm{Cs}^{+}$ and $\rm{Fr}^{+}$ 
respectively. The ratio is much larger in the case $\rm{Fr}^{+}$ and without 
any ambiguity it can be attributed to the relativistic contraction of the 
$np_{1/2}$ orbital. The third leading contribution for $\rm{Na}^{+}$, 
$\rm{K}^{+}$, $\rm{Rb}^{+}$ arise from the $2s_{1/2}$, $3s_{1/2}$  and 
$4s_{1/2}$ orbital respectively. But, for $\rm{Cs}^{+}$ and $\rm{Fr}^{+}$ the 
third leading contribution arise from $4d_{5/2}$ and $5d_{5/2}$ orbital 
respectively. This is because the $5s_{1/2}$ and $6s_{1/2}$ orbital are
contracted due to large relativistic effects. From the above analysis of RPA 
effets, the trend in the contributions demonstrates the importance of 
relativistic corrections in $\rm{Cs}^{+}$ and $\rm{Fr}^{+}$.
\begin{table}[h]
  \caption{Core orbitals contribution from
           $\mathbf{T}_1{^{(1)\dagger}} \mathbf{D}T_2^{(0)}$ to $\alpha$
           of $\rm{Na}^{+}$ and $\rm{K}^{+}$}
    \label{t1dt2_na_k}
    \begin{center}
    \begin{tabular}{dcdc}
       \hline
       \multicolumn{2}{c}{$\rm{Na}^{+}$}  & \multicolumn{2}{c}{$\rm{K}^{+}$} \\
       \hline
       -0.0040  & $(2p_{3/2}, 2p_{3/2})$  &  -0.0646 & $(3p_{3/2}, 3p_{3/2})$ \\
       -0.0021  & $(2p_{3/2}, 2p_{1/2})$  &  -0.0367 & $(3p_{3/2}, 3p_{1/2})$ \\
       -0.0021  & $(2p_{1/2}, 2p_{3/2})$  &  -0.0360 & $(3p_{1/2}, 3p_{3/2})$ \\
       -0.0010  & $(2p_{1/2}, 2p_{1/2})$  &  -0.0148 & $(3p_{1/2}, 3p_{1/2})$ \\
     \hline
    \end{tabular}
    \end{center}
\end{table}

To study the pair-correlation effects, we identify the pairs of core 
spin-orbitals in the next leading order
terms $\mathbf{T}_1{^{(1)\dagger}} \mathbf{D}T_2^{(0)}$  + h.c.
The four leading order pairs for $\rm{Na}^{+}$ and $\rm{K}^{+}$,
$\rm{Rb}^{+}$, $\rm{Cs}^{+}$  and  $\rm{Fr}^{+}$ are listed in table 
\ref{t1dt2_na_k} and \ref{t1dt2_rb_cs_fr} respectively. The dominant
contribution, for all the ions, arise from the combination
$(np_{3/2}, np_{3/2})$ orbital pairing. To illustrate the relative values,
the contributions from the pairs of the five outermost core spin-orbitals
of Rb$^+$ is shown as a barchart in Fig. \ref{rb_chart}.
\begin{table}[h]
  \caption{Core orbitals contribution from
           $\mathbf{T}_1{^{(1)\dagger}} \mathbf{D}T_2^{(0)}$ to $\alpha$
           of $\rm{Rb}^{+}$, $\rm{Cs}^{+}$  and $\rm{Fr}^{+}$}
    \label{t1dt2_rb_cs_fr}
    \begin{center}
    \begin{tabular}{dcdc}
       \hline
       \multicolumn{2}{c}{$\rm{Rb}^{+}$}  & \multicolumn{2}{c}{$\rm{Cs}^{+}$} 
                    \\ \hline
       -0.1113  & $(4p_{3/2}, 4p_{3/2})$  &  -0.2126 & $(5p_{3/2}, 5p_{3/2})$\\
       -0.0601  & $(4p_{3/2}, 4p_{1/2})$  &  -0.1073 & $(5p_{3/2}, 5p_{1/2})$\\
       -0.0565  & $(4p_{1/2}, 4p_{3/2})$  &  -0.0930 & $(5p_{1/2}, 5p_{3/2})$\\ 
       -0.0223  & $(4p_{1/2}, 4p_{1/2})$  &  -0.0347 & $(5p_{1/2}, 5p_{1/2})$\\ 
     \hline
       \multicolumn{2}{c}{$\rm{Fr}^{+}$} && \\ \hline 
       -0.3078 & $(6p_{3/2}, 6p_{3/2})$  && \\
       -0.1266 & $(6p_{3/2}, 6p_{1/2})$  && \\
       -0.0828 & $(6p_{1/2}, 6p_{3/2})$  && \\
       -0.0489 & $(6p_{3/2}, 5d_{5/2})$  && \\ \hline
    \end{tabular}
    \end{center}
\end{table}
Comparing the results of all the ions, there is a major difference in the 
results of Fr$^+$. For Fr$^+$ the fourth largest contribution is from the 
$(6p_{3/2}, 5d_{5/2})$ pair, whereas for the other ions it is of the form
$(np_{1/2}, np_{1/2})$. This is again a consequence of the contraction of 
the $6s_{1/2}$ spin-orbital in Fr$^+$ due to relativistic effects. 

  In the present calculations we have identified the following possible 
sources of uncertainty. The truncation of the spin-orbital basis sets is one 
of the possible source. For all the ions we start the computations with 9 
symmetries and increase up to 13 symmetries. Along with it, we also vary the 
number of the spin-orbitals till $\alpha$ converges to $\approx 10^{-4}$. So, 
we can safely neglect this uncertainty for our calculations. Another
source of uncertainty is the truncation of the CC theory at the single and 
double excitation for both unperturbed and the perturbed RCC theories. 
Based on our previous theoretical results 
\cite{chattopadhyay-12a,chattopadhyay-12b} the contributions from
triple and higher order excitations is at the most $\approx 3.3\%$. The 
truncation of $e^{{\mathbf{T}^{(1)}}^\dagger}\mathbf{D}e^{T^{(0)}} 
+ e^{{T^{(0)}}^\dagger}\mathbf{D}e^{\mathbf{T}^{(1)}}$ at the second order in 
cluster operator is also a source of uncertainty. From our earlier studies
\cite{mani-10} with CC theory and in the present work we have studied the
contribution from the third order in cluster operator, but the contribution
is negligibly small. The quantum electrodynamical(QED) corrections is another
source of uncertainty in our calculations and based on our previous studies,
we estimate it at 0.1\%. In total, we estimate the uncertainties in our 
results as $\approx$3.4\%.


\section{Conclusion}

  We have computed the static electric dipole polarizability of alkali ions
using the PRCC theory. The PRCC theory is a coupled-cluster based theory and 
can be easily modified to incorporate other perturbations in the atomic 
many-body calculations. In the present work, we have explored the use of PRCC 
theory to calculate the electric dipole polarizability of closed-shell ions 
and find that the results are in good agreement with the experimental results 
and previous theoretical results. 

  On a closer examination of the results, the pattern of the contributions 
from the individual and pairs of spin-orbitals establishes the importance
of the relativistic corrections in higher $Z$ ions. The results further
indicates that it is essential to obtain the outermost $p_{3/2}$ spin-orbitals
of the ions accurately. The reason is, these are associated with the dominant 
contributions from the Dirac-Fock, RPA effects and pair-correlation effects.


\begin{acknowledgments}
We thank S. Gautam, Arko Roy and Kuldeep Suthar for useful discussions. The
results presented in the paper are based on the computations using the 3TFLOP
HPC Cluster at Physical Research Laboratory, Ahmedabad.
\end{acknowledgments}

\bibliography{references}{}
\bibliographystyle{apsrev4-1}

\end{document}